\documentclass[11pt]{article}
\usepackage[utf8]{inputenc}
\usepackage{amsmath}
\usepackage{amssymb}
\usepackage{tensor}
\usepackage{mathrsfs}
\usepackage{cite}
\usepackage{multirow}
\usepackage{framed}
\usepackage{graphicx}
\usepackage[colorlinks=true,urlcolor=blue,anchorcolor=blue,citecolor=blue,filecolor=blue,linkcolor=blue,menucolor=blue,linktocpage=true,pdfproducer=medialab,pdfa=true]{hyperref}
\usepackage{calc}
\usepackage[symbol]{footmisc}

\newcommand\bom[1]     {{\mbox{\boldmath $#1$}}}
\newcommand{\La}    {\bom{L_{a}}}
\newcommand{\Lb}    {\bom{L_{b}}}
\newcommand{\Lab}    {\bom{L_{ab}}}

\newcommand{\ILa}    {\bom{I L_{a}}}
\newcommand{\ILb}    {\bom{I L_{b}}}
\newcommand{\ILab}    {\bom{I L_{ab}}}

\makeatletter

\@addtoreset{equation}{section}

\def\draftdate{\relax}
\def\mda{\relax}
\def\mua{\relax}
\def\mla{\relax}
\def\draft{
\def\thtystars{******************************}
\def\sixtystars{\thtystars\thtystasr}
\typeout{}
\typeout{\sixtystars**}
\typeout{* Draft mode!
         For final version remove \protect\draft\space in source file *}
\typeout{\sixtystars**}
\typeout{}
\def\draftdate{\today}
\def\mua{\marginpar[\boldmath\hfil$\uparrow$]%
                   {\boldmath$\uparrow$\hfil}%
                    \typeout{marginpar: $\uparrow$}\ignorespaces}
\def\mda{\marginpar[\boldmath\hfil$\downarrow$]%
                   {\boldmath$\downarrow$\hfil}%
                    \typeout{marginpar: $\downarrow$}\ignorespaces}
\def\mla{\marginpar[\boldmath\hfil$\rightarrow$]%
                   {\boldmath$\leftarrow $\hfil}%
                    \typeout{marginpar: $\leftrightarrow$}\ignorespaces}
\overfullrule 5pt
\oddsidemargin -12mm
\marginparwidth 29mm
}

\def\starline{\hfil\strut\hfil\hbox to \textwidth {\stasr}\hfil}

\def\beq{\begin{equation}}
\def\eeq{\end{equation}}
\def\bsp#1\esp{\begin{split}#1\end{split}}
\def\bal#1\eal{\begin{align}#1\end{align}}

\def\beeq{\begin{eqnarray}}
\def\eeeq{\end{eqnarray}}

\newcommand{\eps}      {\varepsilon}

\newcommand{\rd}       {{\mathrm{d}}}

\newcommand{\cII}[1] {{\cal I}\kern-4pt *\kern-4pt{\cal I}_{#1}}
\newcommand{\cIJ}     {{\cal I}\kern-4pt *\kern-4pt{\cal J}}
\newcommand{\cJJ}[1] {{\cal J}\kern-4pt *\kern-4pt{\cal J}_{#1}}
\newcommand{\cJI}     {{\cal J}\kern-4pt *\kern-4pt{\cal I}}
\newcommand{\cJK}     {{\cal J}\kern-4pt *\kern-4pt{\cal K}}
\newcommand{\cKJ}[1] {{\cal K}\kern-4pt *\kern-4pt{\cal J}_{#1}}
\newcommand{\cKI}     {{\cal K}\kern-4pt *\kern-4pt{\cal I}}
\newcommand{\cKK}     {{\cal K}\kern-4pt *\kern-4pt{\cal K}}

\newcommand{\bSCS}[1]  {\bom{\mathrm C}\kern-2pt\bom{\mathrm S}_{#1}}

\newcommand{\cSCS}[2]  {{\cal C}\kern-2pt{\cal S}_{#1}^{#2}}

\newcommand{\IcSCS}[2]  {\mathrm{C}\kern-2pt\mathrm{S}_{#1}^{#2}}

\def\s12{s_{12}}

\newcommand{\colorful}{CoLoRFul }

\begin{document}

\numberwithin{equation}{section}

\begin{titlepage}
\noindent
DESY-24-190 \hfill December 2024\\
\vspace{0.6cm}
\begin{center}
{\LARGE \bf 
    \colorful for hadron collisions: Integrating the counterterms\footnote[1]{Presented at the V4-HEP 3 - Theory and Experiment in High Energy Physics Workshop, Prague, Czech Republic, 1-4 October 2024.}
}
\vspace{1.0cm}

\large
S.~Van Thurenhout$^{\, a}$\footnote[2]{Speaker}, V. Del Duca$^{\,b}$, C. Duhr$^{\, c}$, L. Fek\'esh\'azy$^{\,d,e}$, F. Guadagni$^{\, f}$, P. Mukherjee$^{\, d}$, G. Somogyi$^{\, a}$ and F. Tramontano$^{\, g}$\\
\vspace{0,5cm}
\normalsize
{\it $^{\, a}$HUN-REN Wigner Research Centre for Physics, Konkoly-Thege Mikl\'os u. 29-33, 1121 Budapest, Hungary}\\
{\it $^{\, b}$INFN, Laboratori Nazionali di Frascati, 00044 Frascati (RM), Italy}\\
{\it $^{\, c}$Bethe Center for Theoretical Physics, Universität Bonn, D-53115, Germany}\\
{\it $^{\, d}$II. Institut für Theoretische Physik, Universität Hamburg, Luruper Chaussee
149, 22761, Hamburg, Germany}\\
{\it $^{\, e}$Institute for Theoretical Physics, ELTE Eötvös Loránd University, Pázmány Péter sétány
1/A, 1117, Budapest, Hungary}\\
{\it $^{\, f}$Physik-Institut, Universität Zürich, 8057 Zürich, Switzerland}\\
{\it $^{\, g}$Dipartimento di Fisica Ettore Pancini, Universit`a di Napoli Federico II and INFN - Sezione di
Napoli, Complesso Universitario di Monte Sant’Angelo Ed. 6, Via Cintia, 80126 Napoli, Italy}
\vspace{1.4cm}

{\large \bf Abstract}
\vspace{-0.2cm}
\end{center}
In order to numerically compute scattering cross sections in QCD, one needs to deal with various kinematic divergences that appear at intermediate stages of the calculation. One way of doing this is by setting up an IR subtraction scheme. In this talk we give an update on the status of extending the \colorful subtraction scheme, which has been successfully used in the past for processes with only final-state hadrons, to hadron-hadron collisions. In particular we discuss the analytic computation of the integrated counterterms.
\vspace*{0.3cm}
\end{titlepage}
\clearpage






\section{Introduction}
\label{sec:Introduction}
Despite the tremendous success of the Standard Model (SM) of particle physics, it is by now well-known that the appearance of new physics at some energy scale is inevitable. Strong hints that lead towards this conclusion include the apparent existence of dark matter and dark energy, the infamous matter-anti-matter asymmetry and the tiny neutrino masses. Unfortunately, no convincing evidence of new physics has been found at high-energy colliders such as the LHC. This, of course, does not necessarily mean it is not there. In particular, beyond the SM physics signals could come to us \textit{indirectly} as small deviations between SM predictions and experimental measurements. This motivates the push towards higher precision of theoretical predictions. In a perturbative quantum field theory context, this of course means computing higher order corrections to physical observables. Such higher order computations are complicated due to the appearance of various kinematic divergences. In particular, problems arise when loop momenta become large (leading to UV singularities) and when momenta become soft and/or collinear to one another (leading to IR singularities). The former are treated once and for all by renormalization. The latter in principle cancel for so-called infrared safe observables~\cite{Sterman:1977wj} when all perturbative contributions are properly taken into account. This is a direct consequence of the Kinoshita–Lee–Nauenberg theorem~\cite{Nakanishi:1958gt,Kinoshita:1962ur,Lee:1964is} and would be the end of the story if one could perform the computation, in particular the phase space integrations, analytically. In practice this is rarely the case however, and one needs to turn to numerical methods. As such, the IR divergences need to be treated explicitly. One way of doing so is by setting up a local subtraction scheme, which entails the construction of an approximate cross section that matches the point-wise singularity structure of the original one. This boils down to a redistribution of the IR singularities leading to separately finite blocks, which can then be evaluated numerically. In the context of strongly interacting particles, the construction of this approximate cross section is guided by factorization and the universal nature of the singularity structure of QCD matrix elements~\cite{Collins:1985ue,Collins:1989gx,Collins:2004nx}. Of course, whatever was subtracted from the cross section to make it IR finite needs to be integrated over the appropriate phase space and added back. These integrations are performed analytically. This way, one can make sure that the subtraction scheme actually works (i.e., that poles cancel analytically). At NLO accuracy, the cancellation of IR divergences by way of local subtractions is considered to be solved~\cite{Catani:1996vz}. The extension to NNLO however is a field of active research~\cite{Catani:2007vq,Gaunt:2015pea,Gehrmann-DeRidder:2005btv,Czakon:2014oma,Caola:2017dug,Cacciari:2015jma,Magnea:2018hab,Herzog:2018ily}. In this proceeding, we give an update on the status of the \colorful framework~\cite{Somogyi:2005xz
}, which so far has been successfully applied to processes with only final-state hadrons. In particular, we discuss the analytic integration of the subtraction terms for hadron-initiated processes. 

This proceeding is organized as follows. In Sec.~\ref{sec:scheme} we briefly review the \colorful subtraction scheme at NNLO accuracy. The next section then provides an overview of the main steps in the analytic computation of the integrated counterterms while Sec.~\ref{sec:NNLOCAL} introduces the {\tt Fortran} implementation of the scheme, dubbed {\tt NNLOCAL}. A brief summary and outlook are presented in Sec.~\ref{sec:summary}.

\section{Review of the subtraction scheme}
\label{sec:scheme}
We consider a hadron collision leading to the production of some colour-singlet state $X$ and $m$ jets. At NNLO, the associated partonic cross section reads
\begin{align}
\label{eq:NNLOxs}
    \sigma^{\text{NNLO}} &= \int_{m+2}\rd \sigma_{m+2}^{\text{RR}}J_{m+2}+\int_{m+1}\rd \sigma_{m+1}^{\text{RV}}J_{m+1}+\int_{m}\rd \sigma^{\text{VV}}J_{m} \nonumber\\& + \int_{m+1}\rd \sigma_{m+1}^{C_1}J_{m+1}+\int_{m}\rd \sigma_{m}^{C_2}J_{m}
\end{align}
in which we explicitly suppress the dependence on partonic momenta and the renormalization and factorization scales. The top line contains the double real (RR), real-virtual (RV) and double-virtual (VV) contributions, while the bottom line incorporates the collinear remnants. The latter only appear for hadron-initiated processes and take into account PDF renormalization. For an infrared safe jet function $J_m$, the full sum in Eq.~(\ref{eq:NNLOxs}) is finite. However, separately the integrals diverge and require regularization. For example, in the \colorful framework the RR cross section is regularized as follows
\begin{align}
\label{eq:NNLORR}
    \sigma_{m+2}^{\text{NNLO}} = \int_{m+2}\Bigg\{&\rd\sigma_{m+2}^{\text{RR}}J_{m+2}-\rd\sigma_{m+2}^{\text{RR},A_1}J_{m+1}-\rd\sigma_{m+2}^{\text{RR},A_2}J_{m}\nonumber\\&+\rd\sigma_{m+2}^{\text{RR},A_{12}}J_{m}\Bigg\}\,.
\end{align}
Per construction, this expression is finite in four dimensions. Each term on the right-hand side of Eq.~(\ref{eq:NNLORR}) is constructed in such a way that it cancels a specific kinematic divergence,
\begin{itemize}
    \item $\rd\sigma_{m+2}^{\text{RR},A_1}$ cancels the singularities coming from a single unresolved emission,
    \item $\rd\sigma_{m+2}^{\text{RR},A_2}$ cancels the singularities coming from a double unresolved emission and
    \item $\rd\sigma_{m+2}^{\text{RR},A_{12}}$ cancels the singularities coming from single (double) unresolved limits in $\rd\sigma_{m+2}^{\text{RR},A_2}$ ($\rd\sigma_{m+2}^{\text{RR},A_1}$).
\end{itemize}
These counterterms now need to be added back, integrated over the appropriate phase space. In this work we focus on the integral of the $A_{12}$ approximate cross section, which needs to be added to Eq.~(\ref{eq:NNLORR}) as
\begin{equation}
\label{eq:A12int}
    -\int_{m}\left[\int_{2}\rd\sigma_{m+2}^{\text{RR},A_{12}}\right]J_{m}.
\end{equation}
In the following, we will concentrate on the production of a colour-singlet final state, i.e. $m=0$.

\section{Integrating the $A_{12}$ subtraction terms}
\label{sec:int}
We are interested in the analytic computation of the integral in Eq.~(\ref{eq:A12int}). The explicit construction of the $A_{12}$ approximate cross section in the colour-singlet case will be discussed in detail in a future publication. Here we simply state that the complete set of $A_{12}$ counterterms leads to 104 basic integrals. These need to be calculated to the appropriate order in the dimensional regulator $\eps$. In this proceeding, we give a generic overview of the steps needed in the integration procedure, leaving the explicit computation for a future publication. In general, the integrated counterterm, which we denote by $\mathcal{IC}$, contains some complicated multidimensional integrals. For example, we often need to compute four-fold integrals of the type
\begin{equation}
\label{eq:IC}
    \mathcal{IC} = \int_{0}^{1}\rd \eta_a\,\int_{0}^{1}\rd \eta_b\,\underbrace{\int_{0}^{1}\rd \xi_a\,\int_{0}^{1}\rd \xi_b\,f(\xi_a,\xi_b,\eta_a,\eta_b;\eps)}_{\mathcal{I}(\eta_a,\eta_b;\eps)}|\mathcal{M}(\eta_a p_a,\eta_b p_b)|^2.
\end{equation}
Here $\mathcal{M}( p_a, p_b)$ is the Born matrix element for the $a(p_a)b(p_b)\to X$ process. The integrand $f(\xi_a,\xi_b,\eta_a,\eta_b;\eps)$ is typically some complicated function of the integration variables and $\eps$. For example, without providing any details, one particular form that we encounter is the following
\beq
\bsp
&f(\xi_a,\xi_b,\eta_a,\eta_b;\eps) = \frac{- (2 - \xi_{a} + \eta_{a} \xi_{a} - \xi_{b} + \eta_{b} \xi_{b})^{-1 + 2 \eps}}{((-1 + \eta_{b}) \xi_{b} + (-1 + \eta_{a}) \xi_{a} (1 + (-1 + \eta_{b}) \xi_{b}))}\\&\times\Big[(\eta_{b} (-1 + \xi_{a})^2 +\eta_{a}^2 \eta_{b} (-1 + \xi_{a}^2) +\eta_{a} (1 - \eta_{b} (\eta_{b} + 2 (-1 + \xi_{a}) \xi_{a}) - 2 \xi_{b}\\& + 2 \eta_{b} \xi_{b} + (-1 + \eta_{b})^2 \xi_{b}^2))
\Big]^{-1}(1 - \eta_{a})^{1 - 2 \eps} \eta_{a}^{-\eps} (1 - \eta_{b})^{-1 - 2 \eps}\eta_{b}^{-\eps} (1 - \xi_{a})^{-\eps}\\&\times \xi_{a}^{-\eps} (1 - \xi_{a} + \eta_{a} \xi_{a})^{-\eps} (2 - \xi_{a} +\eta_{a} \xi_{a})^{-\eps}(1 + \eta_{a} - \xi_{a} + \eta_{a} \xi_{a})^{-\eps} (1 - \xi_{b})^{-1 - \eps}\\&\times\xi_{b}^{-1 - \eps} (1 - \xi_{b} + \eta_{b} \xi_{b})^{ 2 - \eps} (2 - \xi_{b} + \eta_{b} \xi_{b})^{-1 - \eps}(1 + \eta_{b} - \xi_{b} + \eta_{b} \xi_{b})^{-1 - \eps} \\&\times(\eta_{a} + \eta_{b} - \eta_{b} \xi_{a} + \eta_{a} \eta_{b} \xi_{a} - \eta_{a} \xi_{b} + \eta_{a} \eta_{b} \xi_{b})^{-1 + 2 \eps} (2 - (1 - \eta_{b}) \xi_{b} \\&- (1 - \eta_{a}) \xi_{a} (1 - (1 - \eta_{b}) \xi_{b})) ((1 - \xi_{a}) (1 - (1 - \eta_{b}) \xi_{b}) +\eta_{a} (\eta_{b} + \xi_{a} \\&- \xi_{a} \xi_{b} + \eta_{b} \xi_{a} \xi_{b})) (-(1 - \xi_{b}) (\xi_{a} - \xi_{b}) +\eta_{b} \xi_{b} (-1 - \xi_{a} + 2 \xi_{b}) + \eta_{a} (\eta_{b} + \xi_{a} \\&- \xi_{a} \xi_{b} + \eta_{b} \xi_{a} \xi_{b}) -\eta_{b}^2 (-1 + \xi_{b}^2))\,.
\esp
\eeq
The final result of the integration procedure is expected to contain multiple polylogarithms (MPLs)~\cite{Goncharov:1998kja}
\begin{equation}
\label{eq:MPL}
    G(a_1,\dots,a_n;z) = \int_{0}^{z}\frac{\rd t}{t-a_1}G(a_2,\dots,a_n;t)\,, \qquad G(z) \equiv G(;z) =1\,.
\end{equation}
For this reason, we make intensive use of the {\tt PolyLogTools} package~\cite{Duhr:2019tlz}. Assuming we start with the inner integration over $\xi_b$, the main steps to compute the integral in Eq.~(\ref{eq:IC}) can now be summarized as follows:
\begin{enumerate}
    \item Disentangle any overlapping singularities in the integrand using sector decomposition~\cite{Heinrich:2008si}. This way, one obtains a form of the integrand in which all singularities are factorized.
    \item In order to obtain a solution in terms of MPLs, all higher-order polynomials in the denominators of the integrand should be factorized. As generically we have quadratic and quartic polynomials, this factorization will lead to polynomials of the remaining variables with fractional exponents $1/2$ and $1/4$. These should be rationalized, which can be achieved automatically using the {\tt RationalizeRoots} package~\cite{Besier:2019kco}.
    \item Finally, before performing the integration the integrand needs to be partial fractioned in the integration variable. Due to the complexity of our expressions, this turned out to be a major bottleneck. This lead us to develop a new routine for the computation of univariate partial fraction decompositions called {\tt LinApart}~\cite{Chargeishvili:2024nut}. The latter is based on a closed-form expression for the decomposition following from the residue theorem and leads to significant speed-ups with respect to publicly available tools.
\end{enumerate}
After these three steps, one can analytically perform the $\xi_b$-integration using the {\tt GIntegrate} command provided by {\tt PolyLogTools}. The same steps should then be repeated for the $\xi_a$-integration, leading to a complicated expression $\mathcal{I}(\eta_a,\eta_b;\eps)$. Finally, we still need to perform the integration over $\eta_a$ and $\eta_b$,
\begin{equation}
\label{eq:Ietas}
    \mathcal{IC} = \int_{0}^{1}\rd\eta_a\,\int_{0}^{1}\rd\eta_b\,\mathcal{I}(\eta_a,\eta_b;\eps)|\mathcal{M}(\eta_a p_a,\eta_b p_b)|^2.
\end{equation}
One needs to be careful with the interpretation of this integral, as $\mathcal{I}(\eta_a,\eta_b;\eps)$ actually diverges when any of the integration variables approaches one
\begin{equation}
   \mathcal{I}(1,\eta_b;\eps) \rightarrow \infty\,,\qquad\mathcal{I}(\eta_a,1;\eps) \rightarrow \infty\,,\qquad\mathcal{I}(1,1;\eps) \rightarrow \infty\,.
\end{equation}
Because of this, we require an additional regularization, which is accomplished by means of a distributional expansion, i.e. a reinterpretation in terms of delta functions and plus-distributions. This is done by setting up an appropriate subtraction. In particular, we use the method of expansion by regions \cite{Beneke:1997zp} to compute the the asymptotic behaviour of $\mathcal{I}(\eta_a,\eta_b;\eps)$ in all limits with the help of the {\tt asy} package~\cite{Pak:2010pt,Jantzen:2012mw}.
The full integrated counterterm then takes on the form
\beq
\bsp
\mathcal{IC}=&\int_0^1 \rd \eta_a\, \rd \eta_b\,
\bigg\{
[\mathcal{I}(\eta_a,\eta_b;\eps)]|\mathcal{M}(\eta_a p_a,\eta_b p_b)|^2
\\&-
[\La\,\mathcal{I}(\eta_a,\eta_b;\eps)]|\mathcal{M}( p_a,\eta_b p_b)|^2
-
[\Lb\,\mathcal{I}(\eta_a,\eta_b;\eps)]|\mathcal{M}(\eta_a p_a, p_b)|^2
\\&-
\Big([\Lab\,\mathcal{I}(\eta_a,\eta_b;\eps)]-[\La\Lab\,\mathcal{I}(\eta_a,\eta_b;\eps)]\\&-[\Lb\Lab\,\mathcal{I}(\eta_a,\eta_b;\eps)]\Big)|\mathcal{M}(p_a, p_b)|^2+[\ILa\,\mathcal{I}(\eta_b;\eps)]|\mathcal{M}( p_a,\eta_b p_b)|^2\\&+
[\ILb\,\mathcal{I}(\eta_a;\eps)]|\mathcal{M}(\eta_a p_a, p_b)|^2
+\Big([\ILab\,\mathcal{I}(\eps)]-[\ILa\Lab\,\mathcal{I}(\eta_b;\eps)]\\&-[\ILb\Lab\,\mathcal{I}(\eta_a;\eps)]\Big)|\mathcal{M}( p_a,p_b)|^2
 \bigg\}\,
\esp
\eeq
with
\begin{align}
    &\La\,\mathcal{I}(\eta_a,\eta_b;\eps) \equiv \lim_{\eta_a\rightarrow 1} \mathcal{I}(\eta_a,\eta_b;\eps)\,,\\
    &[\ILa\,\mathcal{I}(\eta_b;\eps)] \equiv \int_0^1\rd \eta_a\, [\La\,\mathcal{I}(\eta_a,\eta_b;\eps)]
\end{align}
and similarly for $\Lb, \ILb\,$ etc. This methodology was used for all integrals in $A_{12}$. 


\section{Putting everything together: {\tt NNLOCAL}}
\label{sec:NNLOCAL}
All analytic formulae have been implemented in a {\tt Fortran} code called {\tt NNLOCAL}\footnote{More details on the code were provided during a talk at the HP2 conference by F. Tramontano, \url{https://agenda.infn.it/event/35067/contributions/241516/}.}. As an example of the application of our code, we considered gluon fusion Higgs production in an effective theory in which the top quark is integrated out. Furthermore, for the moment we assume there are no light quarks, i.e. $n_f=0$. After combining the integrals of the $A_{12}$ counterterms discussed in this proceeding with all the other integrated counterterms and the known poles of the collinear factorization, we have verified \textit{analytically} that all poles cancel the ones of the partonic matrix elements at every order in the dimensional parameter $\varepsilon$, proving that our subtraction works as expected. Furthermore, at the inclusive level, we can compare our predictions with those of existing tools, such as {\tt n3loxs}~\cite{Baglio:2022wzu}. For this comparison, we performed calculations for the LHC with 13~TeV center of mass energy using the {\tt NNPDF31\_nnlo\_as\_0118} PDF set~\cite{NNPDF:2014otw}. As highlighted in Table~\ref{tab:xstot}, we find sub-percent agreement for various values of the Higgs mass.
\begin{table}[ht]
\setlength{\tabcolsep}{12pt}
\renewcommand{\arraystretch}{1.5}
    \centering
    \begin{tabular}{|c|c|c|}
    \hline
         $m_H$ [GeV] &  {\tt n3loxs} (gg) &  {\tt
         NNLOCAL} (gg) \\
    \hline\hline
         125 & $42.934$ pb & $42.84 \pm 0.08$ pb\\ 
    \hline
         250 & $9.7290$ pb & $9.717 \pm 0.017$ pb \\ 
    \hline
         500 & $1.6253$ pb & $1.622 \pm 0.003$ pb \\ 
    \hline
         1000 & $173.59$ fb & $173.5 \pm 0.3$ fb \\ 
    \hline
         2000 & $8.7835$ fb & $8.781 \pm 0.017$ fb \\ 
    \hline
    \end{tabular}
    \vspace{1em}
    \caption{Comparison of inclusive $gg\rightarrow H$ cross section predictions at NNLO accuracy between {\tt n3loxs} and {\tt NNLOCAL} for the scale choice $\mu_R = \mu_F = m_H$. The errors on the results obtained with {\tt NNLOCAL} represent the estimated uncertainties of the Monte Carlo integrations while the estimated uncertainty of the {\tt n3loxs} result is beyond the last displayed digit in each case.}
    \label{tab:xstot}
\end{table}

\section{Summary and outlook}
\label{sec:summary}
The application of the CoLoRFul subtraction scheme to colour-singlet production in hadron-hadron collisions at NNLO accuracy is now within reach. In particular, all integrated counterterms have been computed, and the scheme is implemented in the {\tt Fortran} code {\tt NNLOCAL}. For now the latter was used to compute gluon fusion Higgs production in an effective approach, and the extension to full QCD is in progress.

\subsection*{Acknowledgements}
This work has been supported by grant K143451 of the National Research, Development and Innovation Fund in Hungary and the Bolyai Fellowship program of the Hungarian Academy of Sciences. The work of C.D. was funded by the European Union (ERC Consolidator Grant LoCoMotive 101043686). Views and opinions expressed are however those of the author(s) only and do not necessarily reflect those of the European Union or the European Research Council. Neither the European Union nor the granting authority can be held responsible for them. The work of L.F.\ was supported by the German Academic Exchange Service (DAAD) through its Bi-Nationally Supervised Scholarship program.

{\scriptsize
\bibliographystyle{JHEP}
\bibliography{colorful}
}


\end{document}